\documentclass[sigconf,authorversion]{acmart}

\AtBeginDocument{%
  \providecommand\BibTeX{{%
    \normalfont B\kern-0.5em{\scshape i\kern-0.25em b}\kern-0.8em\TeX}}}

\acmConference[PEARC '20]{Practice and Experience in Advanced Research Computing}{July 26--30, 2020}{Portland, OR, USA}

\usepackage{graphicx}
\usepackage{subcaption}
\usepackage{cellspace}

\newcommand{\tabcell}[1]{\begin{tabular}{@{}c@{}}#1\end{tabular}}
\graphicspath{{pics/}}

\begin{document}

\title{Design and Deployment of Photo2Building: A Cloud-based Procedural Modeling Tool as a Service}

\author{Manush Bhatt}
\email{bhatt16@purdue.edu}
\affiliation{%
  \institution{Purdue University}
}

\author{Rajesh Kalyanam}
\email{rkalyana@purdue.edu}
\affiliation{%
  \institution{Purdue University}
}

\author{Gen Nishida}
\email{gnishida@purdue.edu}
\affiliation{%
  \institution{Purdue University}
}

\author{Liu He}
\email{he425@purdue.edu}
\affiliation{%
  \institution{Purdue University}
}

\author{Christopher K May}
\email{may5@purdue.edu}
\affiliation{%
  \institution{Purdue University}
}

\author{Dev Niyogi}
\email{dniyogi@purdue.edu}
\affiliation{%
  \institution{Purdue University}
}

\author{Daniel Aliaga}
\email{aliaga@purdue.edu}
\affiliation{%
  \institution{Purdue University}
}

\renewcommand{\shortauthors}{Manush Bhatt et. al.}

\begin{abstract}

We present a Photo2Building tool to create a plausible 3D model of a building from only a single photograph. Our tool is based on a prior desktop version which, as described in this paper, is converted into a client-server model, with job queuing, web-page support, and support of concurrent usage. The reported cloud-based web-accessible tool can reconstruct a building in 40 seconds on average and costing only 0.60 USD with current pricing. This provides for an extremely scalable and possibly widespread tool for creating building models for use in urban design and planning applications. With the growing impact of rapid urbanization on weather and climate and resource availability, access to such a service is expected to help a wide variety of users such as city planners, urban meteorologists worldwide in the quest to improved prediction of urban weather and designing climate-resilient cities of the future.

\end{abstract}



\begin{teaserfigure}
    \centering
    \begin{subfigure}[h]{0.2\textwidth}
        \includegraphics[width=\textwidth]{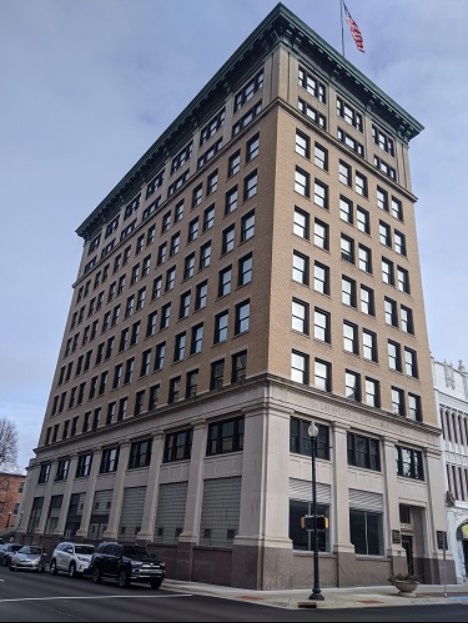}
        \caption{Input Image}
    \end{subfigure}
    \begin{subfigure}[h]{0.2\textwidth}
        \includegraphics[width=\textwidth]{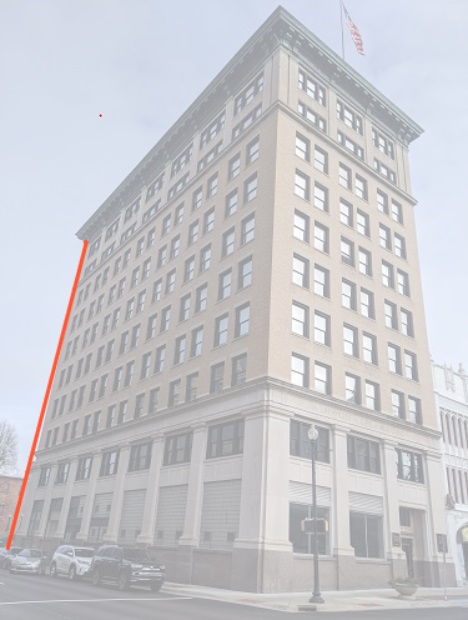}
        \caption{Begin by drawing a line}
    \end{subfigure}
    \begin{subfigure}[h]{0.2\textwidth}
        \includegraphics[width=\textwidth]{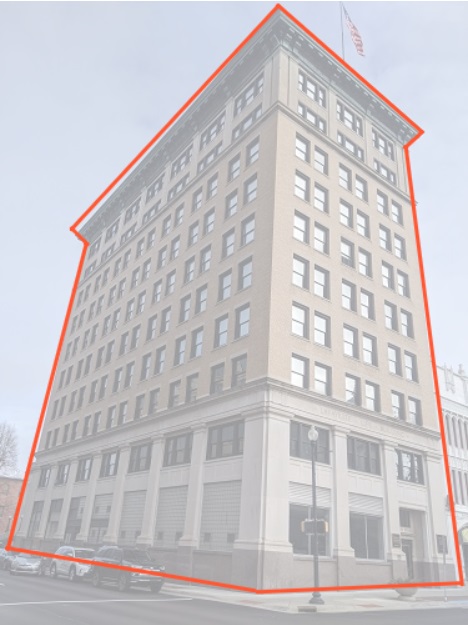}
        \caption{Complete Silhouette}
    \end{subfigure}
    \begin{subfigure}[h]{0.143\textwidth}
        \includegraphics[width=\textwidth]{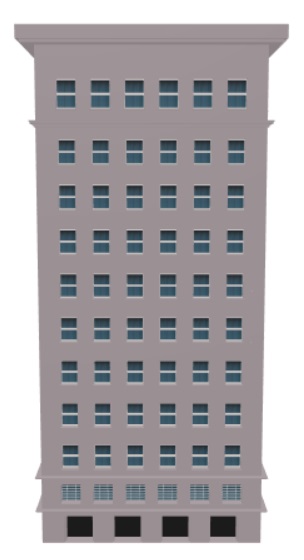}
        \caption{3D Output Front}
    \end{subfigure}
    \begin{subfigure}[h]{0.225\textwidth}
        \includegraphics[width=\textwidth]{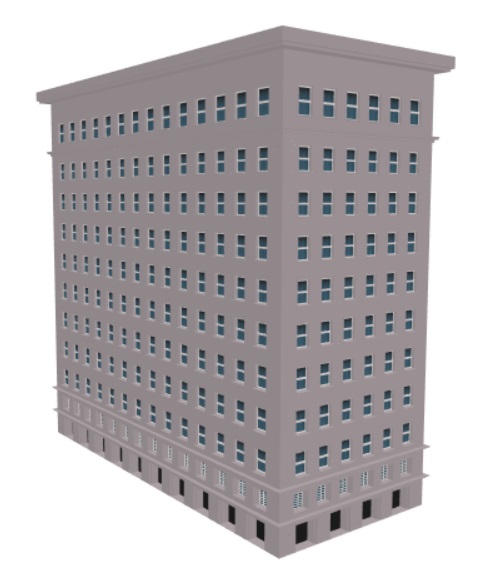}
        \caption{3D Output Side}
    \end{subfigure}
    \caption{Photo2Building: a web-based interactive tool for automatic 3D building modeling}
\end{teaserfigure}

\maketitle

\section{Introduction} \label{introduction}

\textbf Urbanization is a fast-emerging, global phenomenon. With the increase in urban population and infrastructure demands, the need to design and plan the evolution as well as the creation of city components is of critical importance. Even though cities only occupy 3\% of Earth's surface, they have a significant and a disproportionately large effect on the environment and human life. Thus modeling, simulating, and developing scenarios related to future cities is only continuing to increase in importance.

\textbf Concurrently, advances in the development of secure, robust, and highly scalable cyberinfrastructure have enabled ease of public access to high-performance computing, data repositories, and seamless research collaboration. Cloud infrastructure providers such as Amazon Web Services (AWS), Google Cloud Platform (GCP), and Microsoft Azure are increasingly being incorporated in modern cyberinfrastructure environments. For example, Yang et al. \cite{cyberinfrastructureGEE} used Google Earth Engine (GEE) along with Google Cloud to establish a highly-scalable and high-performance analysis cyberinfrastructure. We seek to leverage both the computational capacity of cloud computing and modern cyberinfrastructure platforms to deploy a scalable, publicly accessible service that leads to improvements in urban design, if-then scenario building for environmental models, and planning.

\textbf Recent developments in deep learning technologies, using GPUs, can provide functionalities previously demanding massive computer power or considered infeasible for non-specialized applications. Our driving inspiration in undertaking this work is to create a novel tool that exploits deep learning and cloud-based GPUs to provide the broad ability to produce an entire 3D building model from a single photograph. This approach, rather than requiring significant detailed information and numerous pictures of a single building, enables rapidly creating a plausible model of a building and then making it available for urban modeling and planning tasks. This methodology can easily be used globally due to cloud computing and storage, and the computation time and costs are expected to be extremely low, thus enabling widespread deployment.

\textbf Our approach consists of a web-based client with an interactively-controlled tool, a server "in the cloud" able to process the compute jobs from a few to a large number of simultaneous users, and a deep-neural-network back-end system. The tool allows a user anywhere in the world to upload a photograph of a building and provides back in under a minute a plausible 3D model of the building, in a common object file format. Our tool makes uses of procedural modeling, which is a process of creating 3D models based on a predefined set of rules. This technique is used for generating complex models such as terrains, buildings, plants, and cities. It provides flexibility to represent 3D models by varying a set of parameterized grammars. However, creating grammar rules for virtual architectures from scratch is difficult and time-consuming.  The online cloud-based tool we provide is based on our prior work \cite{Nishida2018ProceduralMO}, which automatically generates procedural grammar from a single image by combining parameter generation with machine learning. We evaluate our system with more than 100 images taken from the ImageNet database \cite{imagenet_cvpr09}, Sun database, and from Lafayette, IN. We report computing times and storage needed, as well as cost estimates. In summary, it costs about 0.60 USD and 40 seconds of computing time to reconstruct a single building, which is significantly lower than prior efforts.

This paper is organized as follows. Section \ref{background} describes background information and related work. Section \ref{design} describes the overall design of the tool as a service. Sections \ref{implementation_client}, \ref{implementation_server}, and \ref{implementation_building_rec} gives a detailed view of the underlying implementation. The results are described in Section \ref{results} and conclusions and discussion about ongoing and future work are presented in Section \ref{conclusion}.

\section{Background} \label{background}

Musialski et al. \cite{Musialski} provide a broad summary of urban reconstruction work, including the wide range of practical benefits from the reconstructed urban models. Vergauwen et al. \cite{VanGool3DRec} describes a web-based 3D rendering service for rendering a scene or an object of cultural heritage. The 3D reconstruction is developed from multi-view images using GPU programming. Urban reconstruction is also achieved using photogrammetry, which is a process of extracting 3D models of an object or terrain from a set of 2D images.  For example, Ruan et al. \cite{Photogrammetry} offer a photogrammetry service with similar goals to our tool. 

One component of urban modeling is procedural methods that have been used in the construction of complex environments such as terrains, animations, and objects (e.g. Parish and Mueller \cite{ParishMueller}). While procedural modeling is widely used in the ESRI's CityEngine tool, to our knowledge, ours is the first service to provide an automatic procedural modeling tool from a single image.








MyGeoHub~\cite{mygeohub} is used to host a publicly accessible web-based tool. MyGeoHub is a publicly accessible science gateway built on the HUBzero cyberinfrastructure (CI) framework~\cite{hubzero}. HUBzero provides several capabilities for online, collaborative science. These include the ability to manage, share, and publish data files and resources such as presentations, tools, and course materials. Tools on HUBzero-based platforms can span a wide variety of programming languages and can be accessed and run directly from a web browser. HUBzero also recently added support for building tools based on the Jupyter and R Studio interactive computing environments. All HUBzero tools can access high performance and cloud computing resources via a job submission system called ``submit'' that can be used to submit jobs, transfer files, and retrieve outputs from these remote resources. The details of the implementation of this MyGeoHub tool is described in Sections \ref{design} and \ref{implementation_client}.

\section{Design} \label{design}

To provide a web-accessible interface to our Photo2Building tool, we decouple the neural network computation that utilizes GPUs from the user interface of the tool. We conceptualize this decoupling as a client-server architecture where the server performs the necessary GPU computation (on cloud computing or some other HPC resource provider), and the client provides the user interface for managing the inputs and visualizing the outputs of the procedural modeling. 

\begin{figure}[]
    \centering
    \frame{\includegraphics[width=\linewidth, keepaspectratio]{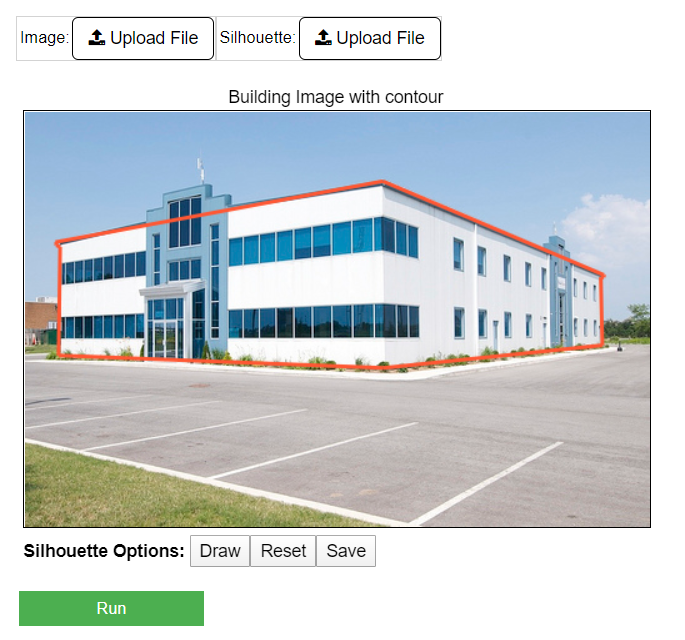}}
    \caption{Client GUI Interface}
    \label{fig:client_gui}
\end{figure}

We envisage that the role of the client is to provide a graphical user interface and connect with a job management system to send and receive data with the server. HUBzero's ``submit'' serves as this job management system that connects the client to the server. Figure \ref{fig:client_gui} depicts a typical view of the system's GUI in a web browser. The interface allows the user to upload an image of a building, draw a silhouette demarcating the building, upload a pre-existing silhouette, or save the current silhouette to local disk. The interface can also view a rendered 3D model of the automatically reconstructed building using WebGL. The steps necessary to use the GUI are outlined in figures \ref{fig:gui_01} and \ref{fig:gui_03}.





The role of the server is to perform the actual job of building reconstruction and to return the output to the client. A server is a virtual machine hosted on Google Cloud with an attached GPU, and the necessary NVIDIA CUDA drivers installed. The server code runs in containers for ease of deployment, portability, and scalability. A simple job queue is utilized on the server to handle concurrent user requests. Figure \ref{fig:server_arch} shows the overall architecture of the server.

\begin{figure}[H]
    \centering
    \includegraphics[width=\linewidth, keepaspectratio]{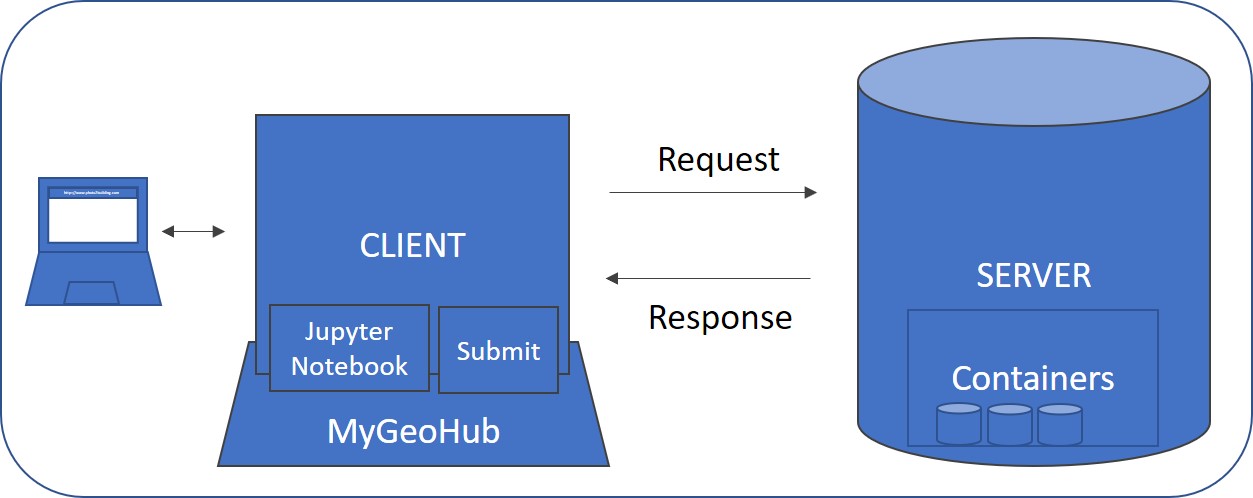}
    \caption{Client-server model of Photo2Building Tool}%
    \label{fig:design}
\end{figure}

The building-reconstruction per se is performed by processing the user-provided silhouette of the building and the photograph of the building through a multi-stage pipeline that combines image processing algorithms, optimization, and deep neural networks. The end-output is a 3D building model in the Wavefront OBJ file (a well-known object format for 3D models).

\section{Client} \label{implementation_client}

The Photo2Building client application is accessed by signing in through MyGeoHub and launching the Photo2Building Jupyter tool. Our initial goal was to separate the original desktop tool code into the server and client with minimal changes. In this original design, the client would be a Linux desktop tool based on Qt5 and OpenGL, utilizing MyGeoHub's support for running Linux desktop tools in OpenVZ containers accessible via VNC remote desktop infrastructure. We found several limitations with this approach. First, the MyGeoHub tool container operating system did not support the Qt and OpenGL versions used by the desktop tool. Our attempts to use WebGL instead of OpenGL had its downsides. This approach would only work for OpenGL (Embedded System) calls and was incompatible with the Qt5 Widgets module (used by our original tool) and other non-OpenGL requests. Hence, we decided to develop the client GUI from scratch using a WebGL based Javascript library, Jupyter notebooks and interactive ipywidgets, and HTML5.

\subsection{Tool Usage}

\begin{figure}[]
    \centering
    \includegraphics[width=\linewidth, ]{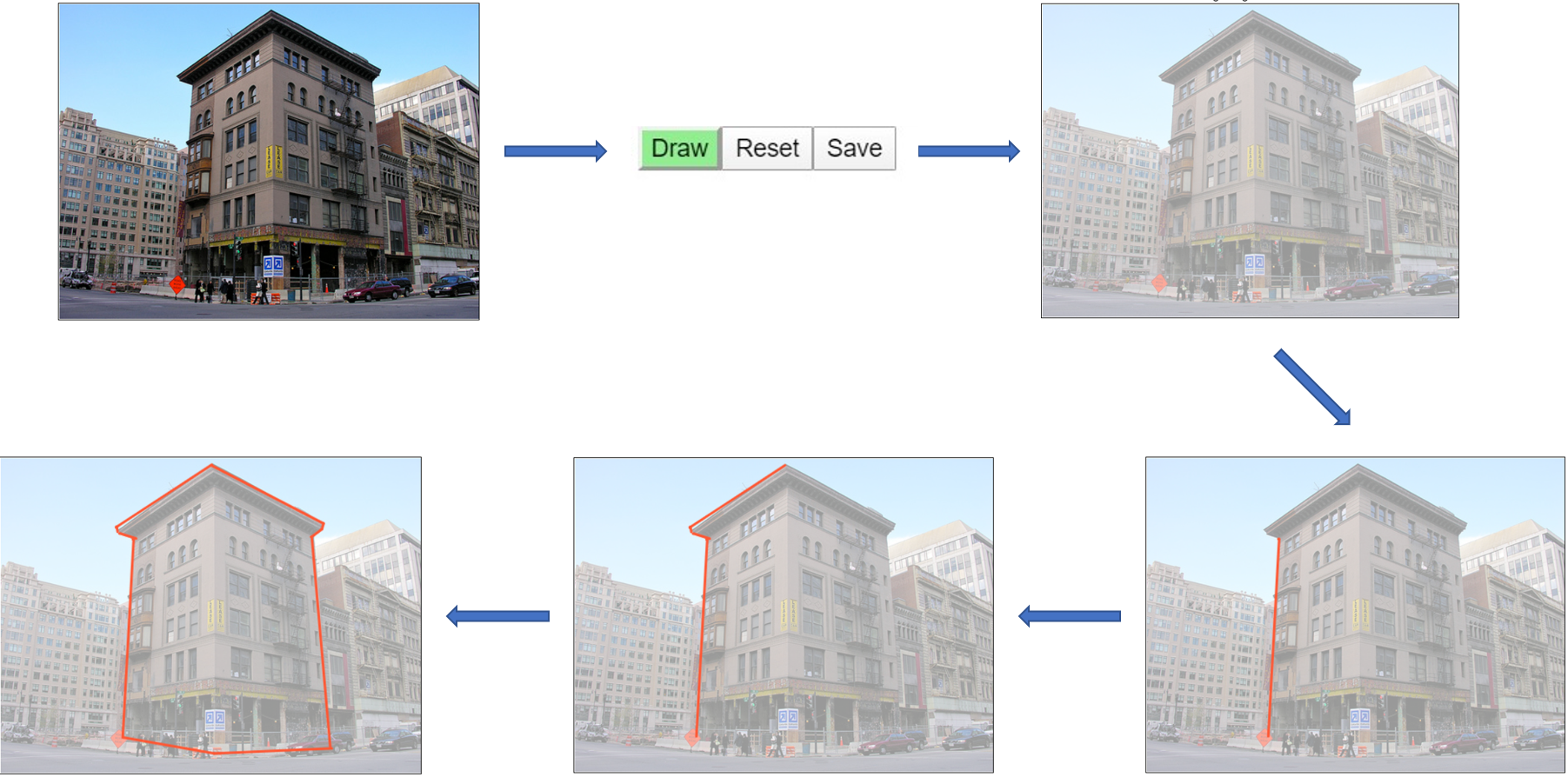}
    \caption{Procedure for drawing a silhouette}
    a) Upload an input image. b) Enable draw mode. c) Draw a line along the building boundary. d) Plot additional lines using the previous as the reference and form the curvature. e) Complete the silhouette for the entire building.
    \label{fig:gui_01}
\end{figure}

\begin{figure}[]
    \centering
    \frame{\includegraphics[width=0.5\linewidth]{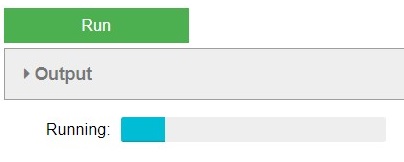}}
    \caption{Run Job and Report Progress}%
    {To run the 3D building reconstruction, the user clicks on "Run" and can track the status and progress of the job}
    \label{fig:gui_03}
\end{figure}

The user starts by uploading a photograph containing the target building. On upload, the "Draw" mode is enabled, allowing the user to draw a silhouette around the building using the cursor. The client also provides options to perform a "Reset" as well as to "Save" the silhouette in the form of a text file containing coordinate information of the various line segments comprising the silhouette. The saved silhouettes can be re-used along with the corresponding building photograph in subsequent uses. The various pieces of the client GUI are illustrated in Figures \ref{fig:gui_01} and \ref{fig:gui_03}.

When the user is satisfied with the silhouette of the target building, the job for building model generation is initialized with "Run". Behind the scenes, HUBzero ``submit'' is used to package and send the input data (building image and silhouette files) to the server and to execute a job script on the server for processing this input data. On job completion, the server returns the output in the form of an object (".obj") file along with the required textures and materials for the walls, windows, and roof. The detailed implementation of server and building reconstruction are described in \ref{implementation_server} and \ref{implementation_building_rec}. The returned object file is rendered on the client GUI through the use of the Three.js Javascript 3D library. The user can freely interact with the rendered 3D model as well as download the object and material files. These files can be used with popular 3D graphics desktop tools such as Blender and MeshLab.

\section{Server} \label{implementation_server}
The server is a more complex part of the infrastructure deployed using Google Cloud Compute Engine. It uses a containerized ecosystem to isolate the process of building reconstruction as well as perform jobs in a concurrent fashion. When a new job is submitted from the client, HUBzero ``submit'' executes a job script on the server that triggers the processing of the input files for this particular job and waits until the job completes execution or a pre-determined timeout of 2 minutes is encountered. On the exit of this job script, ``submit'' automatically returns any output files to the client.

\begin{figure}[]
    \centering
    \includegraphics[width=\linewidth]{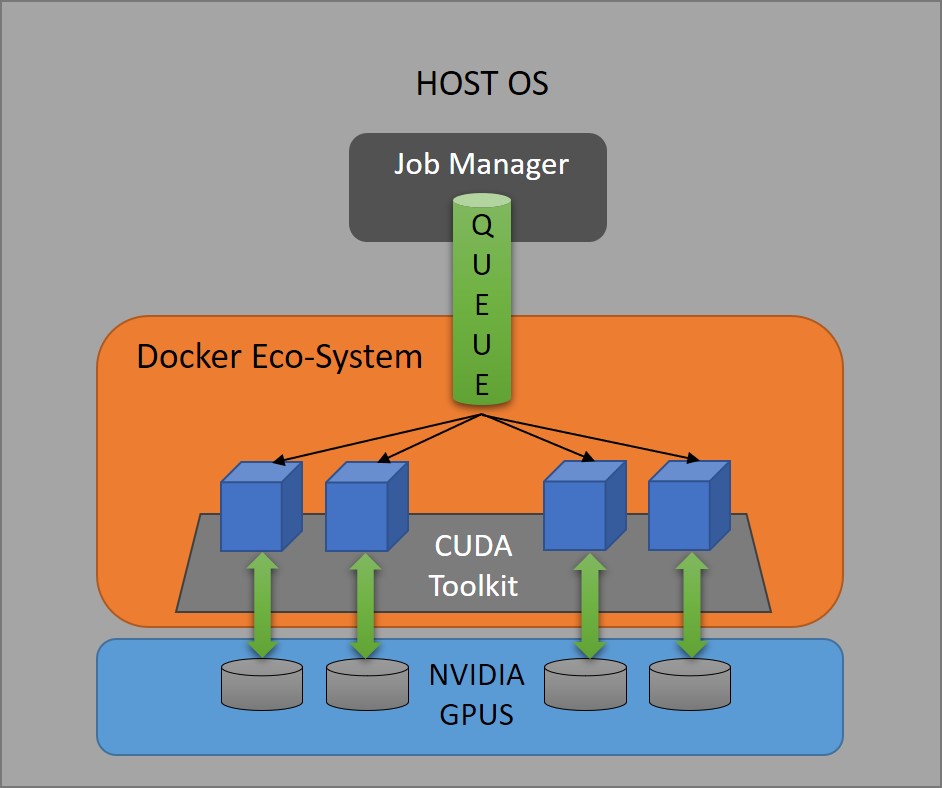}
    \caption{Server Architecture}
    \label{fig:server_arch}
\end{figure}

We use the in-memory data store, Redis, to implement a simple job queue to handle concurrent requests. As the first step in the job script, a message in JSON format is inserted into a Redis queue with the specifics of the input files and auto-incremented job ID generated by ``Submit'' for this job. The server containers access this Redis queue to determine the next job to be executed. Since reads are ``blocking'', exactly one container will process a given message. On completion of the processing, the container sets the job's status in a hash structure mapped with the job's ID. The job script periodically polls this hash structure to determine if a particular job has completed. Figure \ref{fig:server_arch} shows an in-depth architecture of the server-side implementation.

The server containers use the Docker containerization framework and are built on base images provided by NVIDIA. The NVIDIA CUDA Deep Neural Network cuDNN library is a GPU acceleration library for applications using deep neural networks. An Ubuntu 18.04 base layer containing the cuDNN library \cite{NvidiaDockerImage} is used as the base layer for assembling the Photo2building Docker layer. We extend this base layer with the necessary libraries, configuration files, CNN models pre-trained using Caffe and the Photo2building application code. The ``nvidia-docker'' NVIDIA container toolkit is used to build and execute this Photo2Building Docker container with GPU acceleration.

The Photo2building application code loads the neural network into memory and waits for job requests indefinitely. On discovering a non-empty queue, it pops the job request from the tail end of the queue. The job request is parsed using a JSON decoder and the process of building reconstruction is initiated with the decoded input parameters. The process of building reconstruction is described in section \ref{implementation_building_rec}.

\subsection{Challenges Encountered}

While migrating the original desktop version of the tool to a containerized environment in the cloud, the lack of a destination display surface (buffer) to render the image proved to be an issue. Our tool automatically estimates the camera parameters of the provided image. To do so, it repeatedly optimizes these parameters to minimize the distance between the input and target silhouette by rendering intermediary images on a destination surface buffer. Although, Qt5 supports functionality to run in an off-screen environment, configuring that alone is not enough. Qt5's OpenGL module, which is responsible for rendering the intermediate images, requires a destination surface buffer. EGL provides an interface between various OpenGL APIs and the underlying display window system. It handles the graphics context management as well as the bindings for surface buffer. EGL is supported by NVIDIA drivers and acts as a medium to create rendering surfaces where client APIs like OpenGL are allowed to draw and create graphics contexts. We solve the problem by creating a graphics context using EGL to handle all the OpenGL calls.

Another challenge was the design for sharing resources in a multi-GPU configuration. Our initial approach focused on sharing GPU resources equally between the Docker containers. In that case, the job picked up by a container used any available GPU with free memory. Consequently, the deep neural network model had to be loaded in the GPU memory for each job before reconstructing the 3D model. This loading resulted in extra time for computation spent on reloading the model. To reduce this lead time for each job, we decided to pre-assign GPUs to containers and always kept the model loaded in the GPU's memory. Strictly binding each container to a single GPU also helped in acquiring consistent results from across all containers by avoiding the possibility of multiple containers scrambling for the same resource.


\section{Building Reconstruction} \label{implementation_building_rec}
The process of building reconstruction involves estimating camera parameters and generating grammars that yields a 3D geometry as an output. The process is as described in \cite{Nishida2018ProceduralMO}. There are three steps involved in this process, using recognition Convolutional Neural Networks (CNNs) and parameter estimation CNNs. Recognition CNNs select the procedural grammar rules, and their appropriate parameters are estimated by using parameter estimation CNNs. An overview of each of the three steps is described next. 

\subsection{Building-mass generation}
The first step is responsible for simultaneously determining the building mass style and camera parameters. This simultaneous determination has proven to be much more robust than performing them separately. To determine the building mass style and camera parameters, first a recognition CNN is used. Then, our approach uses a per-style parameter estimation CNN to estimate the actual grammar parameters. Finally, a distance is calculated between the input silhouette $S$ and the generated silhouette $\hat{S}$. The distance is minimized using a  bound-optimization-by-quadratic-approximation (BOBYQA) \cite{bobyqa} algorithm to refine the parameter values.

Each parameter estimation CNN is trained by using 300,000 synthetically created facades. From all these synthetic facades, a 300,000 subset is chosen to train the recognition CNN from many different camera poses. 

\subsection{Facade generation} \label{Facade_gen}
The next step is the initial rectification of the visible facades of the building to get the best representation of the original facades. The rectified facades are then merged to form a single rectified image. The next steps involve facade simplification, refinement, grammar selection, and coloring.

The number of floors and columns of windows is determined by training another CNN using images from CMP facade dataset \cite{FacadeDataSet}. Further, the size of the dataset is increased by performing randomized translations, horizontal mirroring, blurring, and changes to luminosity. The generated grid is then refined using image gradients. The location of windows or doors is discovered using a window recognition CNN from each tile. A window parameter estimation CNN is used to estimate the 2D relative coordinates of the top-left corner of the window/door and the width and height of the window.

A facade recognition CNN is used to recognize the suitable grammar style from a set of 16 defined styles. As with the building-mass grammar styles, each facade grammar style contains multiple parameters such as floor height, window/door size, and appearance, which are estimated using a parameter estimation CNN. Similar to building-mass parameter estimation refinement, a BOBYQA algorithm is used to refine the estimated facade parameters.

The facade color is estimated using a K-Means clustering algorithm on the pixels excluding the detected windows. The centroid of the largest cluster is chosen to be the facade color. The size of the cluster was set to k=10 and the color clustering is performed in L*A*B color space.

\subsection{Window generation}
The estimated facade grammar is decomposed into windows and door tiles. A window/door grammar is selected out of 31 defined grammar styles using a window style recognition CNN. The CMP Facade Dataset, previously defined in Section \ref{Facade_gen}, is also used to generate training images for window tiles. The total number of images generated were 100,000 including augmentations.

The final output combines building-mass grammar, facade grammar and window/door grammars with their respective parameter values. The evaluation images were taken from ImageNet \cite{imagenet_cvpr09} and SUN \cite{SUNDataset}. 

The trained CNN models are saved to local storage. At run-time, the container loads all the models into GPU memory sequentially before serving jobs. This approach saves time otherwise spent on loading the model for each job.


\section{Results and Discussion} \label{results}
\newcolumntype{L}{>{\centering\arraybackslash}p{1cm}}
\begin{figure*}[htbp]
    \centering
    \begin{tabular}{|Sc|Sc|c|c|c|c|Sc|Sc|c|c|c|c|}
        \hline
        Input & Output & Floors & \multicolumn{3}{c|}{Time (In secs)} &  Input & Output & Floors & \multicolumn{3}{c|}{Time (In secs)} \\
        \cline{4-6} \cline{10-12}
        & & & \multicolumn{2}{c|}{Compute} & Total & & & & \multicolumn{2}{c|}{Compute} & Total \\
        \cline{4-5} \cline{10-11}
        & & & Per Floor & Total & & & & & Per Floor & Total & \\ \hline
        \tabcell{\includegraphics[scale=0.14]{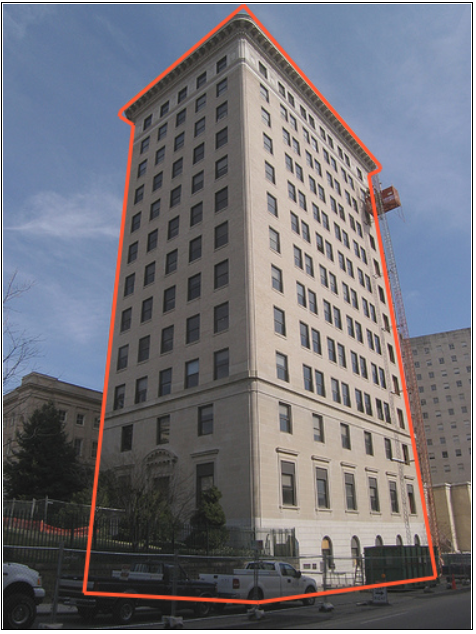}} & \tabcell{\includegraphics[scale=0.14]{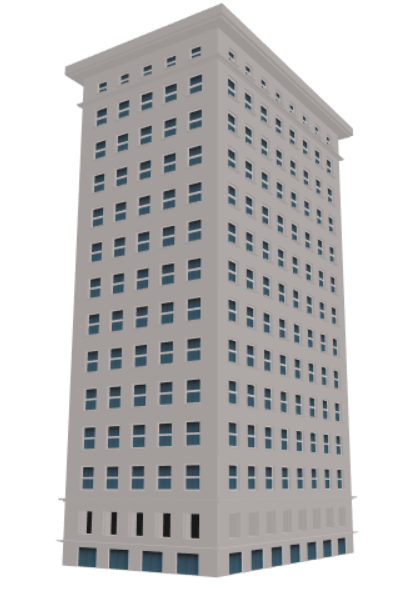}} & 14 & 1 & 14.2 & 38 &
        \tabcell{\includegraphics[scale=0.14]{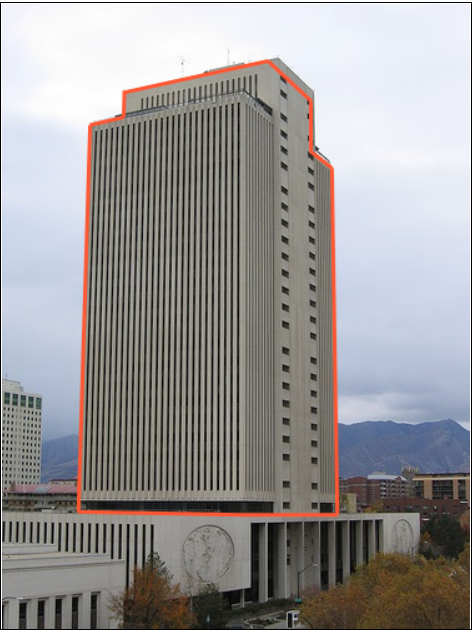}} & \tabcell{\includegraphics[scale=0.14]{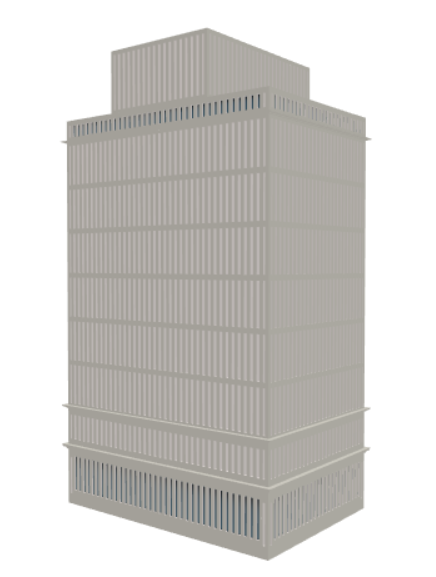}} & 9 & 2 & 17.8 & 41.5  \\ \hline
        \tabcell{\includegraphics[scale=0.14]{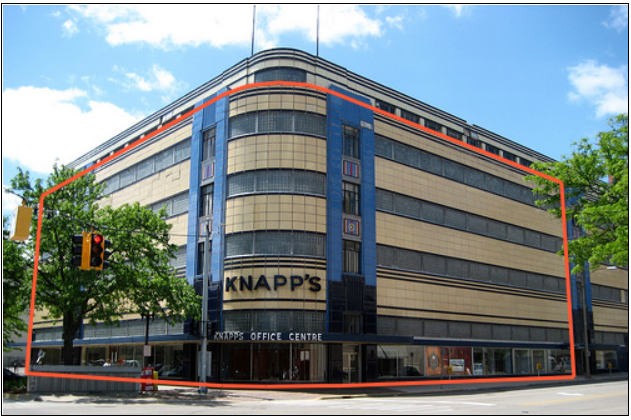}} & \tabcell{\includegraphics[scale=0.14]{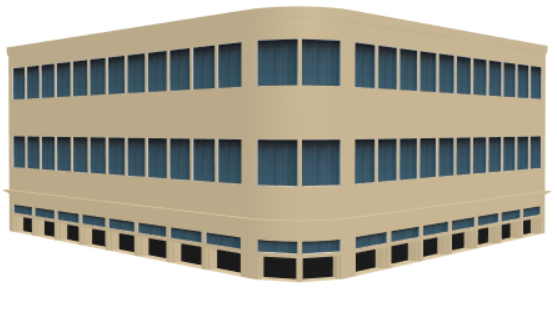}} & 3 & 4.5 & 13.5 & 36.8 & \tabcell{\includegraphics[scale=0.14]{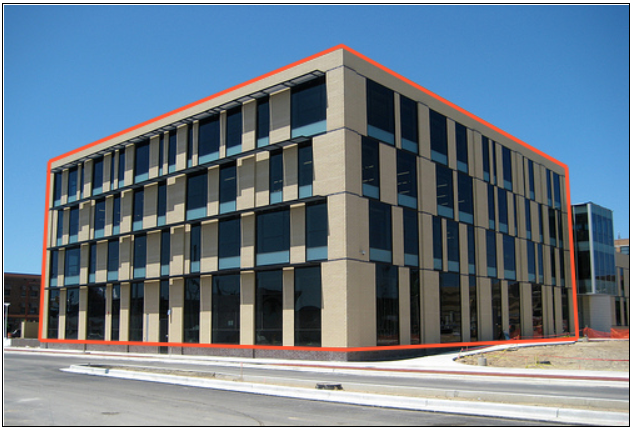}} & \tabcell{\includegraphics[scale=0.14]{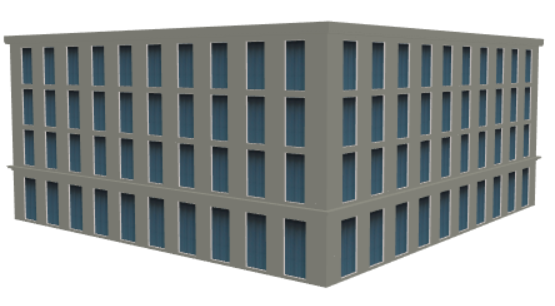}} & 4 & 3.6 & 14.3 & 38 \\ \hline
        \tabcell{\includegraphics[scale=0.14]{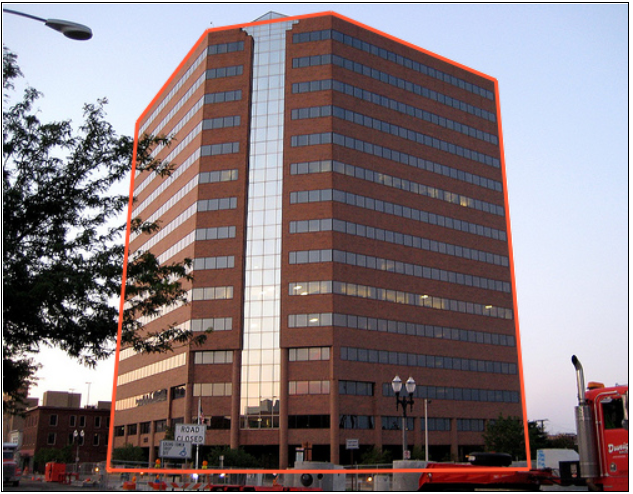}} & \tabcell{\includegraphics[scale=0.14]{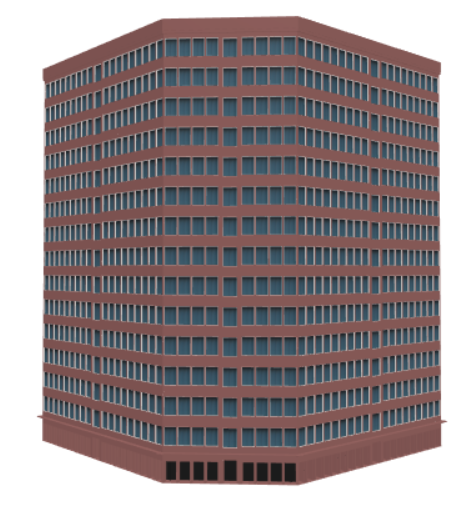}} & 15 & 1.5 & 22.6 & 46.8 & \tabcell{\includegraphics[scale=0.14]{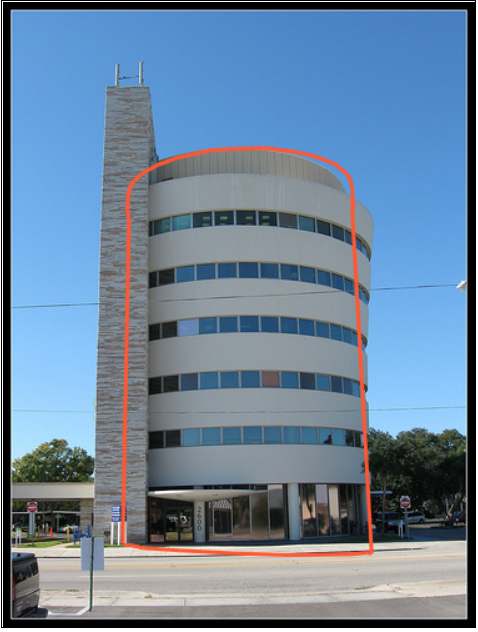}} & \tabcell{\includegraphics[scale=0.18]{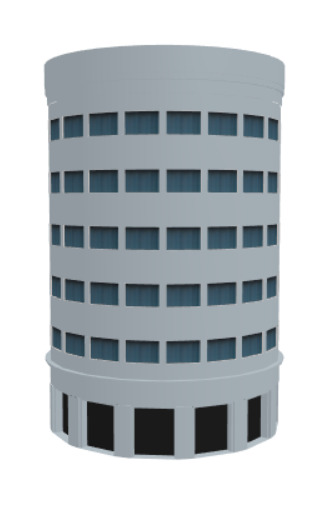}} & 6 & 2.1 & 12.8 & 36.2 \\ \hline
        \tabcell{\includegraphics[scale=0.14]{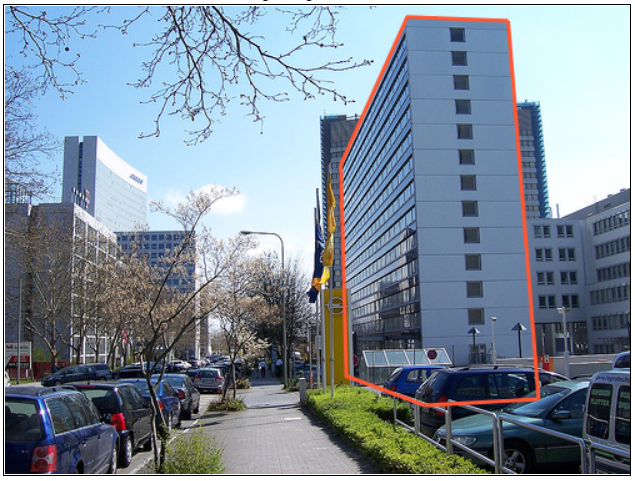}} & \tabcell{\includegraphics[scale=0.18]{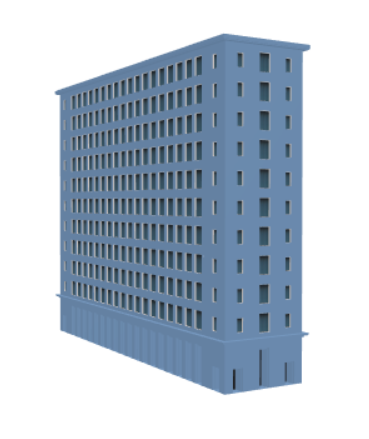}} & 11 & 0.8 & 9 & 32.8 & \tabcell{\includegraphics[scale=0.14]{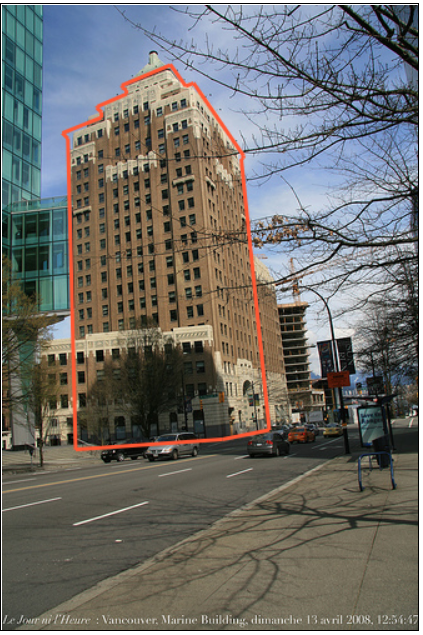}} & \tabcell{\includegraphics[scale=0.18]{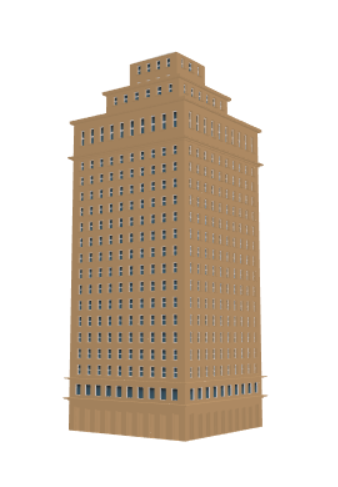}} & 17 & 1.3 & 22 & 46.5 \\ \hline
    \end{tabular}
    \raggedright \textbf{\textit{Average Reconstruction Time per Building: 15.8 sec}} \\
    \textbf{\textit{Average Total Time per Building: 39.6 sec}} \par
    \caption{A comparison of different types of 3D generated buildings. Total time includes time spent on network transfers, submit job scheduling and computations for reconstructing the 3D model. }
    \label{fig:results}
\end{figure*}

We used four NVIDIA Tesla K80 GPUs with eight Intel Xeon CPUs on Google Cloud for our initial evaluation. We received Google Cloud credits for this project through a NSF-Internet2 Exploring Clouds for Acceleration of Science (E-CAS) grant. 

With each client request using a GPU exclusively, this setup was capable of handling four concurrent job requests, while the other requests were queued and executed as jobs were completed. The resolution of the input images used was 512*512. We discovered that the overall time taken for the modeling task was proportional to the complexity of the building structure; i.e., the number of floors and columns. Figure \ref{fig:results} shows a brief comparison of the average processing times for different types of buildings. The average compute and total times were calculated based on processing of 6 samples for each building image.

Even though, our solution supports various building types, the results for some buildings may not be accurate. The solution tries to make the best fit of building-mass, facade and window grammar styles. Along with this, there is a possibility of inaccuracies in the number of floors or columns due to indistinguishable vertical and horizontal edges,  obstruction caused by environmental objects or weather conditions. 

Next, we perform a preliminary cost analysis for the best case scenario of reconstructing a 3D building on Google Cloud. The server is hosted on Google Cloud 24x7 with the aforementioned configuration. The total cost of running the setup is approximately 1200 USD per month. Considering an average time of 60 seconds to construct a single building and taking into account the concurrent nature of our infrastructure to support four simultaneous building reconstructions, in a month up to 172,800 building models can be generated at this price point. Thus, each dollar spent on Google Cloud will effectively create 157 3D buildings. This analysis is pertinent to generating building models at the city scale, discussed as future work.


\section{Conclusions and Future Work} \label{conclusion}

In this paper, we describe a publicly accessible, cloud-based service that can reconstruct 3D buildings efficiently using procedural modeling. The service will generate a complete 3D procedural model of a building given just a single photograph of the building. We utilize a client-server architecture where the client is a web application deployed to a publicly accessible cyberinfrastructure platform, and the server is a scalable, containerized solution that can be deployed to various commercial cloud and HPC resource providers. 

Our work benefits the urban environmental modeling and urban planning community as well as various smart cities' efforts, including World Urban Database and Application Portal Technology (WUDAPT) \cite{CHING2019100459} and other grassroots efforts underway. For instance, in the urban weather/environmental modeling work, a look-up table typically provides information related to the height of the building. The building height and shape information is then used to compute morphological parameters such as aerodynamic roughness and the displacement height, which are used in the dynamical computations related to canyon wind flow, and a thermal changes within the urban canyon. While urban mapping in terms of the impervious surface area and the urban typology is becoming increasingly available through WUDAPT-like efforts, the information about building heights and morphology is still lacking. The Photo2Building tool provides a ready digitization, and development of a dataset that can be populated for such urban environmental models (as an option beyond what is available in the default look-up tables) and is extremely scalable in terms of compute time and cost, as well as widespread availability. 

As part of future work, we will tackle various technical and non-technical challenges:
\begin{itemize}
    \item We are currently exploring the use of Kubernetes~\cite{k8s} to simplify the server container orchestration process as well as perform automatic load balancing and scaling in response to increasing user requests.
    \item The client functionality will be significantly enhanced to allow users to simply specify an input city grid of interest, following which 3D models of all buildings in that city grid will be generated and provided to the user.
    \item The reconstruction code will be enhanced to support for input pictures with varying resolutions. 
    \item The impact of using Photo2Building generated building typology and morphology in an urban environmental (weather) simulation will be undertaken. 
\end{itemize}

\begin{acks}

This research was funded in part by an Internet2 and National Science Foundation ECAS grant \textit{Building Clouds: Worldwide Building Typology Modeling from Images}, NSF \#10001387, \textit{Functional Proceduralization of 3D Geometric Models}, and NSF grant \#1835739, \textit{U-Cube: A Cyberinfrastructure for Unified and Ubiquitous Urban Canopy Parameterization}.

\end{acks}


\bibliographystyle{ACM-Reference-Format}
\bibliography{photo2building}

\end{document}